\documentclass[aip,jmp,reprint,onecolumn]{revtex4-1}
\usepackage{amsfonts}
\usepackage{amsmath}
\usepackage{graphicx}
\usepackage{bm}

\input{tcilatex}

\begin{document}

\title[RSLC in QW]{Resonant Subband Landau Level Coupling in Symmetric
Quantum Well}
\author{L.-C. Tung}
\email{tung@magnet.fsu.edu}
\affiliation{National High Magnetic Laboratory, Tallahassee, Florida 32310}
\author{X.-G. Wu}
\affiliation{Department of Physics, Institute of Semiconductor, Chinese Academy of
Science, China}
\author{L. N. Pfeiffer}
\affiliation{Department of Electrical Engineering, Princeton University, Princeton, New
Jersey 08544, USA}
\author{K. W. West}
\affiliation{Department of Electrical Engineering, Princeton University, Princeton, New
Jersey 08544, USA}
\author{Y.-J. Wang}
\altaffiliation[Note: ]{Dr. Wang passed away in Dec. of 2009 due to cardiac arrest.}
\affiliation{National High Magnetic Laboratory, Tallahassee, Florida 32310}
\date{\today}

\begin{abstract}
Subband structure and depolarization shifts in an ultra-high mobility GaAs/Al%
$_{0.24}$Ga$_{0.76}$As quantum well are studied using magneto-infrared
spectroscopy via resonant subband Landau level coupling. Resonant couplings
between the 1st and up to the 4th subbands are identified by well-separated
anti-level-crossing split resonance, while the hy-lying subbands were
identified by the cyclotron resonance linewidth broadening in the
literature. In addition, a forbidden intersubband transition (1st to 3rd)
has been observed. With the precise determination of the subband structure,
we find that the depolarization shift can be well described by the
semiclassical slab plasma model, and the possible origins for the forbidden
transition are discussed.
\end{abstract}

\maketitle


Many terahertz (THz) radiation applications\cite{Fer02,Sie02,Guo09} involve
detecting or generating THz radiation using intersubband (ISB) transitions
in a quasi-two dimensional electron system (2DES). Designing the 2D
heterostructure optimized for desired THz applications demands a
comprehensive understanding of the subband structure and ISB couplings. Many
intended applications involves optical absorptions in their operations,
which can be affected by the depolarization of the radiation. Subband
structure and depolarization effect were extensively studied in the past,
but the understanding of these subjects are more qualitative than
quantitative and sometimes controversial. It is believed that depolarization
shifts the absorption energies of the ISB transitions, but the magnitude of
the depolarization shift has been calculated using various methods. The
depolarization shift can be calculated numerically in the self-consistent
calculation by calculating induced changes of the charge densities, but it
is not always available and sometimes impossible given the circumstance of
the system.

In the past, two of the analytical models have been developed for
calculating the depolarization shifts. A semiclassical approach is to
calculate the depolarization shift by approximating the 2DES as a slab
plasma of thickness $d$ filled with electrons.\cite{Che76,And82} The
depolarization-shifted ISB transition energy $\tilde{\omega}_{10}$ is
related to the classical plasma frequency $\omega _{p}$ and ISB transition
energy $\omega _{10}$ as $\tilde{\omega}_{10}^{2}=\omega _{10}^{2}+\tilde{%
\omega}_{p}^{2}$, where the effective plasma frequency $\tilde{\omega}_{p}$=$%
(f_{10}\omega _{p}^{2})^{1/2}$ and $f_{10}$ is the oscillator strength of
the optical transition. On the other hand, the depolarization shift was
described by a microscopic model, in which the overlap of the subband
wavefunctions has to be calculated. The shifted ISB transition energy is
represented as $\tilde{\omega}_{10}=\omega _{10}\sqrt{1+\alpha _{10}+\beta
_{10}}$,\cite{And77,Fis83} \ where $\beta _{10}$ represents the exchange
interaction. The depolarization shift is represented by $\alpha _{10}$ and $%
\alpha _{10}$ is related to the overlap of the subband wavefunction $\phi
_{i}(z)$ via $S_{11}=\int_{-\infty }^{\infty }dz[\int_{-\infty
}^{z}dz^{\prime }\phi _{1}(z^{\prime })\phi _{0}(z^{\prime })]^{2}$. \ Both
of the models have been used to calculate the depolarization shift and agree
fairly well with the experimental results. Due to the uncertainty in the
function form of the subband wavefunctions in many 2DES systems, it is not
yet clear whether the depolarization shift is better described by the
microscopic model that considers the wavefunction overlaps, or the
semiclassical model that approximates the 2DES as a three dimensional plasma.

To evaluate the magnitude of the depolarization shift, one should first
measure the ISB transition energies. ISB transitions have been investigated
using optical intersubband-resonance (ISR),\cite{Wie88,And88}
magneto-transport measurements\cite{Ens92,Ens93} and mostly resonant subband
Landau level coupling (RSLC). RSLC measures the ISB transition energies by
coupling the in-plane cyclotron resonance (CR) orbital motion to the motion
along the confinement axes with the presence of a magnetic field parallel to
the 2DES. When the CR energy is brought close to the ISB transition energy,
CR splits into two modes due to the resonant anti-level crossing of the
Landau levels (LLs) belonging to different subbands. It was used to
investigate subband structure in heterojunctions,\cite%
{Sch83,Rik86,Bru86,Wie87,Pil89,Ens89,Mic95,Che97,Sug03} parabolic quantum
wells,\cite{Kar89,Bre89} and (asymmetric) quantum wells.(QW)\cite%
{Sta91,Win96,Orr09} All of these works focused on the ISB transition between
the two lowest subbands and the hy-lying subbands were identified by CR
linewidth broadening.\cite{Mic95,Che97}

A symmetric QW eases the uncertainties arising from the gradient of the
confining potential,\cite{And82} the presence of the depletion charges,\cite%
{Ens89} and subband's diamagnetic shifts and offsets in $k$-space.\cite%
{Bru86} We have selected an ultra-clean, $500\dot{A}$ symmetrically doped Al$%
_{0.24}$Ga$_{0.76}$As/GaAs/Al$_{0.24}$Ga$_{0.76}$As QW ($n_{s}=1.1\times
10^{11}{cm}^{-2}$) and investigated its subband energies via RSLC by
magneto-infrared (IR) spectroscopy at tilt angles from $10^{\circ }$ to $%
35^{\circ }$. In the symmetric QW, the subband wavefunction are better
known, and thus the depolarization shift can be estimated more accurately
using both of the analytical models. We find that the depolarization shift
can be better described by the semiclassical slab plasma model and have
observed a resonant coupling forbidden in the first order perturbation.

A set of magneto-IR spectra ($\theta =25^{\circ }$) are displayed in Fig. 1
(b)-(d) and a schematic energy diagram is displayed in Fig. 1 (a). It is
plotted in scale using the subband energies obtained from the
self-consistent calculation, which excludes the depolarization shift. Since
the 2DES enters the extreme quantum limit at around 2.2T, only the n=0 and
n=1 LLs of the first subband need to be considered. At the resonance of the
LL and ISB transitions, i.e. when the CR energy matches the ISB transition
energy, split resonances result from the anti-level crossing between the n=1
LL of the first subband and the n=0 LLs of the higher subbands. With
increasing magnetic field, the lower-energy mode transfers its integrated
intensity to the higher-energy mode, as shown in Fig. 1 (b)-(d). RSLC to the
hy-lying subbands are observed by the well-separated split resonance and the
energies of the split resonance can be precisely extracted to study the ISB
transitions to the hy-lying subbands. 
\begin{figure}[tbp]
{{{%
\includegraphics[
natheight=11in,
natwidth=8.5in,
height=4in,
width=3.4in
]{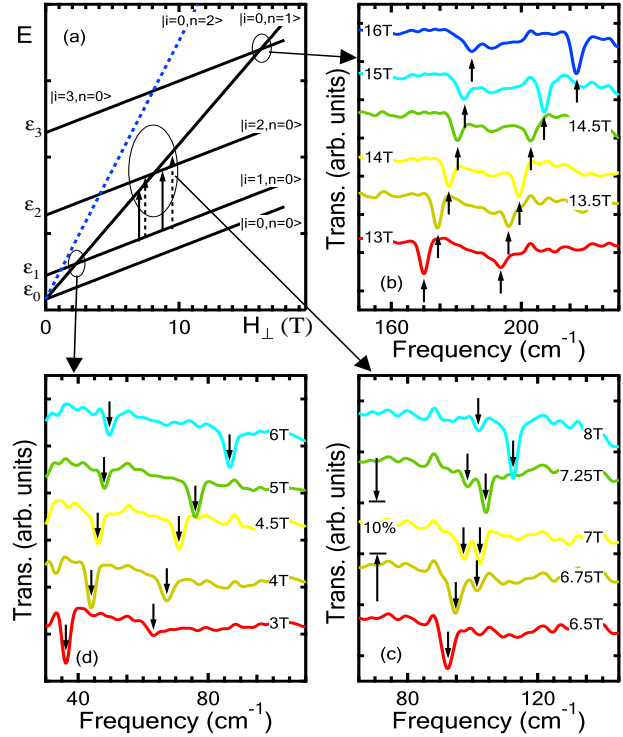}}}}
\caption{(a): A schematic energy diagram for the LLs of the subbands. The
energies of the LLs are shown as a function of the vertical component of the
magnetic field. Each energy level is labeled by the subband index $i$ and LL
index $n$. The anticrossing behavior occurs when the LL transition energy
matches the ISB transition energy, i.e. at the crossing points of the n=1 LL
of the 1st subband and n=0 LL of the higher subbands. (the regions enclosed
by the circles) Around each anticrossing, split resonance results from the
resonant coupling of the LLs belonging to different subbands. The
transitions for the lower branch are shown in solid arrows and the ones for
the higher branch are shown in dashed arrows. Half-field crossings occur at
where the n=2 LL of the 1st subband (shown in blue dotted line) crosses the
n=0 LLs of the higher subbands. One can easily deduce that it will show a
bundle of three transitions, and negative magnetic-field dispersion, i.e.
the transition energy appears to decrease with increasing magnetic field.
(b)-(d): The magneto-infrared spectra for $\protect\theta =25^{\text{o}}$.
The traces are shifted vertically for clarity. (b) $B=13$ to $16{T}$, for
the RSLC between the 1st and 4th subbands. (c) $B=6.5$ to $8{T}$, between
the 1st and 3rd subbands. (d) $B=3$ to 6 T, between the 1st and 2nd
subbands. }
\end{figure}

The energies of the split resonance as a function of the magnetic fields at
five different angles are plotted in Fig. 2. The dotted lines show the
expected CR energies that scale with $\cos \theta $. We will refer the
anticrossings as the first, second and third, ordered by their energies in
ascending order. The energies of the split resonance around the first
anticrossing can be well described by the coupled oscillator model\cite%
{Bor87,Mer87} with $m^{\ast }=0.069m_{e}$, and $\tilde{\omega}_{10}=57
cm^{-1}$.

To get a picture of the transitions particularly for those resulting from
the resonant coupling to the hy-lying subbands, coupled Schr\"{o}dinger and
Poisson equations are solved self-consistently in order to obtain the
subband energy levels in the presence and absence of the magnetic field.\cite%
{And82} The 2DES is confined in a $500\unit{%
\text{\AA}%
}$ QW with finite barrier. The barrier height is determined from the Al
fraction in the barrier material. We have selected the lower one ($180$meV)
of the two band offsets used in the literature. Since the QW is wide, the
lowest subband energy is insensitive to the selection of the barrier height,
while the higher subbands are slightly affected. The conduction-valence
bandgap is around $1520$meV and the magnetic field range in this work is not
too high, so the conduction band non-parabolicity is ignored. Without using
any perturbation, the subband energies and the matrix element for the
depolarization are calculated by an exact diagonalization of the
Hamiltonian, which includes the LL subband coupling.

Using the linear-response approximation, depolarization shifts are taken
into account by considering the induced oscillating electron density along
the direction of the confinement, when the frequency-dependent dynamic
conductivity is evaluated.\cite{And82} The electron density along the well
direction is assumed to be symmetric about the center of the QW. The
formulation is similar to, but not exactly the same as ref. [5]. Our
calculation stops at just calculating the resonance frequencies, whereas the
oscillator strength were also calculated in the literature.

In the numerical calculation, material parameters are taken for GaAs, and
typically $25$ subbands with $8$ or $16$ LLs are included in the self
consistent calculation. In considering the depolarization shift, the lowest $%
20$ energy levels are taken into account. The excitonic shift is ignored,
since its effect is negligible in a wide quantum well.\cite{And82,Kar89} The
result of the self-consistent calculation is shown in solid lines in Fig. 2,
and the calculated transition energies agree well with the measured
transition energies when the depolarization shift are considered. 
\begin{figure}[tp]
{{{{%
\includegraphics[
natheight=11in,
natwidth=8.5in,
height=4.2in,
width=3.4in
]{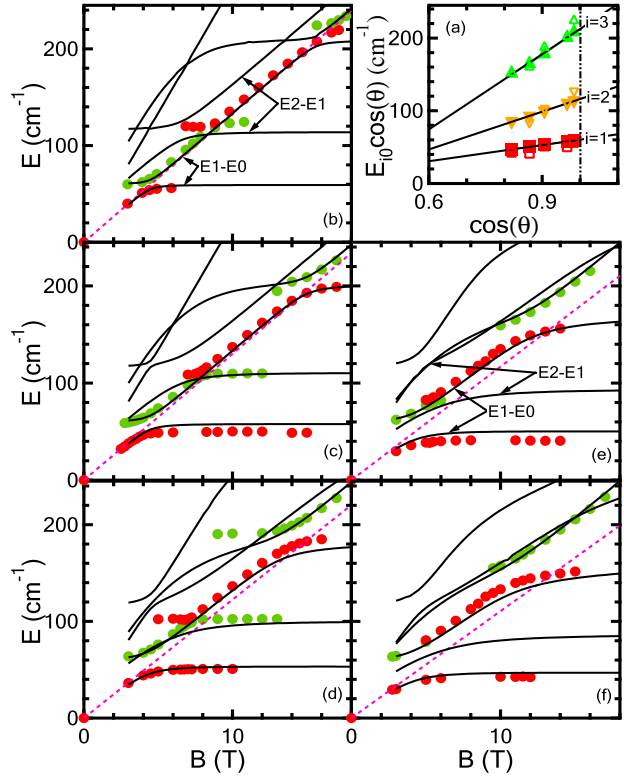}}}}}
\caption{The energies of the split resonance as a function of magnetic field
for different angles: (b) 10$^{\text{o}}$, (c) 15$^{\text{o}}$, (d) 25$^{%
\text{o}}$, (e) 30$^{\text{o}}$ and (f) 35$^{\text{o}}$. The solid lines are
results of the self-consistent calculation including the depolarization
shift. The dotted lines represent the expected CR energies that scale with $%
\cos \protect\theta $. (a): Pinning energies of the lower branches of the
anticrossings are plotted against $\cos \protect\theta $. The pinning
energies extracted from the measurements are shown in open symbols, while
the ones extracted from the self-consistent calculation are shown in solid
symbols. }
\end{figure}

To determine the magnitude of the depolarization shift, we will have to
extract the ISB transition energies from the experimental results and
compared them with the ones calculated by excluding the depolarization
shift. Instead of using the mid-points between the split resonance,\cite%
{Ens89} we extract the ISB transition energies using the pinning energies of
the lower branches of each anticrossing. The lower branches of each
anticrossing pin at around $\tilde{\omega}_{i0}\cos \theta $ at high fields
as shown in Fig. 2 (a). The pinning energies decrease linearly with
decreasing $\cos \theta $ for measured and calculated transition energies.
It is surprising that the linear dependence still holds even when the sample
is tilted $35^{\circ }$. The ISB transition energies can then be extracted
by extrapolating the pinning energy to $\cos \theta \sim 1$ (i.e. $\theta =0$%
) and the results are listed in Table 1. 
\begin{table}[tbp]
\begin{tabular}{|p{0.75in}|l|l|l|}
\hline
& Exp. (${cm}^{-1}$) & Theo. (depol.) & \ ex. depol. \\ \hline
$\tilde{\omega}_{10}$ & \multicolumn{1}{|r|}{$57$} & \multicolumn{1}{|r|}{$%
60 $} & \multicolumn{1}{|c|}{$31$} \\ \hline
$\tilde{\omega}_{20}$ & \multicolumn{1}{|r|}{$116$} & \multicolumn{1}{|r|}{$%
116$} & \multicolumn{1}{|c|}{$110$} \\ \hline
$\tilde{\omega}_{30}$ & \multicolumn{1}{|r|}{$216$} & \multicolumn{1}{|r|}{$%
212$} & \multicolumn{1}{|c|}{$218$} \\ \hline
\end{tabular}%
\caption{Measured subband energies: One column lists the experimental values
for the anticrossings, while the other list the theoretical values when the
depolarization effect is included (depol.) or excluded (ex. depol.).}
\end{table}

By comparing the ISB transition energies in Table 1, the energy of the first
anticrossing is nearly doubled due to the depolarization shift, leaving
others unaltered. Unlike the previous works, we are dealing with a much
simpler system, which leaves narrow margins to fine tune the result. In the
past, studies over the depolarization shifts were carried out mostly on
heterojunctions, in which the gradient of the confining potential and the
presence of the depletion charges were usually unknown, leading to
uncertainties in the wavefunction forms and thus the magnitude of the
depolarization shift. In this wide and symmetric QW, subband wavefunctions
will be close to the ones in the infinite quantum well of the same width.
Using the subband wavefunctions of an infinitely deep QW, the depolarization
shift for the 1st anticrossing can be represented as $\tilde{\omega}%
_{10}^{2}=\omega _{10}^{2}+\frac{5}{3}\omega _{p}^{2}$. Using sample's
parameters, the depolarization shifted ISB transition energy is then $69{cm}%
^{-1}$, but it is much larger than the measured values. One should note that
this discrepancy cannot be overcame by tuning the energy spacings between
the subbands using different material parameters. The measured ISB
transition energy is simply smaller than $\sqrt{\frac{5}{3}}\omega _{p}$,
leaving no space for fine tuning. A finite QW is expected to have an even
larger depolarization shift, since it has been demonstrated by Fishman\cite%
{Fis83} that $S_{11}$ calculated using wavefunctions for a finite QW is
larger than the ones using wavefunctions for an infinite QW. The
depolarization shift for the two hy-lying subbands are negligible, since the
energy spacings are much larger than the plasma frequency $\omega _{p}$ and
the leading factors are small.\cite{Fis83}

Alternatively and more simply, depolarization shifts were calculated using
the slab plasma model.\cite{Che76,And82} Using the oscillator strength for
an infinite QW,\cite{Kar89} $f_{10}\sim 0.96$, the depolarization-shifted
ISB transition energy $\tilde{\omega}_{10}$ is $56{cm}^{-1}$, consistent
with the result of this work. For the 2nd and 3rd anticrossings,
depolarization shifts are minimal, since $f_{30}\sim 0.03$ and $f_{20}\sim 0$
(forbidden). It appears that the results are better described by the
semiclassical model. The proof of this claim can be pursued by finding the
carrier concentration dependence of the depolarization-shifted ISB
transition energies.

The ISB transition between the 1st and 3rd subband is forbidden due to
symmetry, but a resonant coupling between the CR and the forbidden ISB
transition has been observed. With increasing tilt angles, the lower branch
of the anticrossing (E2-E1) between the 3rd and 2nd subbands is depressed
below the upper branch of the one (E1-E0) between the 2nd and the 1st
subbands, thus the 2nd anticrossing becomes more difficult to resolve with
increasing tilt angle. It is not a half-field crossing of the third
anticrossing, since those should be associated with a negative
magnetic-field dispersion and appears as a bundle of three transitions.\cite%
{Wie87} Moreover, half-field crossings require the QW to be asymmetric in
order to have a significant energy separation between the split resonance,%
\cite{Zal89} and sufficient population in the $n=1$ LL of the first subband
to yield sufficient intensity for the half-field-crossing split resonance.
Introducing a modest linear potential along $z$-axes increases the intensity
of this symmetry-forbidden split resonance, but it also significantly
reduces the energy of the third anticrossing.

Exceptions of the selection rules for optical transitions in 2DES have been
reported recently in InSb\cite{Orr09} and InAs\cite{War96} quantum wells,
which were explained in terms of the multiband \textbf{k}$\cdot $\textbf{p}
perturbation theory. In the one band model (conduction band only), a
transition from the 1st to the 3rd subband is not allowed; however, if the
influence of the valence band is considered, the selection rules may be
relaxed,\cite{Orr09} though it will be very weak, since the
conduction-valence bandgap is rather large in GaAs.

Another possibilities is that the ISB transition may still be forbidden, but
a resonant coupling to the forbidden transition becomes possible when the
higher-order couplings are considered. Neglected higher-order terms in the
theoretical calculation\cite{Zal89} may be responsible for this
symmetry-forbidden anticrossing. It has been suggested that QWs may have
large third-order optical nonlinearity\cite{Zal92,Yue83} when the photon
energy matches the ISB transition energy. Some nonlinearity contributions%
\cite{Yue83} depend on a relaxation time, which measures the time needed for
the system to relax from the non-equilibrium state to the equilibrium state.
In this ultra-clean system, it is likely to take longer time for the
electron subsystem to relax, since the scattering rate should be much lower
due to high mobility.

In summary, we have investigated the subband structures and the ISB
transitions via RSLC in an ultra-clean symmetric QW. For the first time,
RSLC to the hy-lying subbands are observed by well-separated split
resonance, including a symmetry-forbidden resonant coupling. We find that
the depolarization shifts can be better described by the slab plasma model.

\begin{acknowledgments}
We like to thank P. Cadden-Zimansky and Jinbo Qi for their assistance. The
measurements were performed at NHMFL at Tallahassee, supported by National
Science Foundation and the state of Florida.
\end{acknowledgments}

\end{document}